\documentstyle[sprocl,epsfig]{article}

\def\Journal#1#2#3#4{{#1} {\bf #2}, #3 (#4)}

\def\NPA{{\em Nucl. Phys.} A}
\def\PLB{{\em Phys. Lett.}  B}
\def\PRL{\em Phys. Rev. Lett.}
\def\PRD{{\em Phys. Rev.} D}

\def\be{\begin{equation}}
\def\ee{\end{equation}}
\def\bea{\begin{eqnarray}}
\def\eea{\end{eqnarray}}

\newcommand{\br}{{\bf r}}

\newcommand{\btab}{\begin{tabbing}\color{green}}
\newcommand{\etab}{\end{tabbing}}
\newcommand{\btablec}{\begin{table}\color{green} \begin{center}}
\newcommand{\etablec}{\end{center} \end{table}}

\newcommand{\beqn}{\footnotesize\begin{equation}\color{dcyan}}
\newcommand{\eeqn}{\end{equation}}

\newcommand{\barr}[1]{\begin{array}{#1}}
\newcommand{\earr}{\end{array}}
\newcommand{\beqna}{\begin{eqnarray}\color{red}}
\newcommand{\eeqna}{\end{eqnarray}}

\newcommand{\gapproxeq}{\lower.7ex\hbox{$\;\stackrel{\textstyle>}{\sim}\;$}}

\begin{document}

\begin{flushright}
nucl-th/0204031 \\
LA-UR-02-2040
\end{flushright}
\vspace{0.6cm}

\title{HYBRID BARYONS
\footnote{Invited plenary talk presented at the ``$9^{th}$ International
Conference on the Structure of Baryons'' (BARYONS 2002), 3--8 March, 
Newport News, VA, USA.}
}

\author{P. R. PAGE}

\address{Theoretical Division, MS B283, Los Alamos National
Laboratory,\\ Los Alamos, NM 87545, USA\\E-mail: prp@lanl.gov} 


\maketitle\abstracts{We review the status of hybrid baryons.
The only known way to study hybrids rigorously is via
excited adiabatic potentials. Hybrids can be modelled
by both the bag and flux--tube models. The low--lying hybrid
baryon is $N\frac{1}{2}^+$ with a mass of $1.5-1.8$ GeV. Hybrid
baryons can be produced in the glue--rich processes of
diffractive ${\gamma} N$ and ${\pi} N$ production, $\Psi$ decays 
and $p\bar{p}$ annihilation.}

\section{Introduction}

We review the current status of research on three quarks
with a gluonic excitation, called a {\it hybrid baryon}.
The excitation is {\it not} an orbital or radial excitation 
between the quarks.
Hybrid baryons have also been reviewed elsewhere.~\cite{bar00}

The Mercedes-Benz logo in Fig.~\ref{merc} indicates two possible
views of the confining interaction of three quarks, an essential 
issue in the study of
hybrid baryons. In the logo the three points where the $Y$--shape
meets the boundary circle should be identified with
the three quarks. There are two possibilities for the interaction 
of the quarks: (1) a pairwise interaction of the quarks
represented by the circle, or (2) a $Y$--shaped interaction
between the quarks, represented by the $Y$--shape in the
logo.

\section{Why does one consider hybrid baryons?}

\hspace{.65cm}(1) {\it You cannot avoid them}. This is because 
excited glue is
predicted by QCD (see the lattice QCD hybrid meson excited adiabatic 
potentials later),
so that hybrid baryon degrees of freedom should be part of
baryon spectroscopy.

\begin{figure}[t]
\begin{center}
\epsfig{file=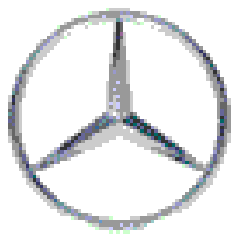,width=.54\linewidth,clip=}
\hspace{-.6cm}\epsfig{file=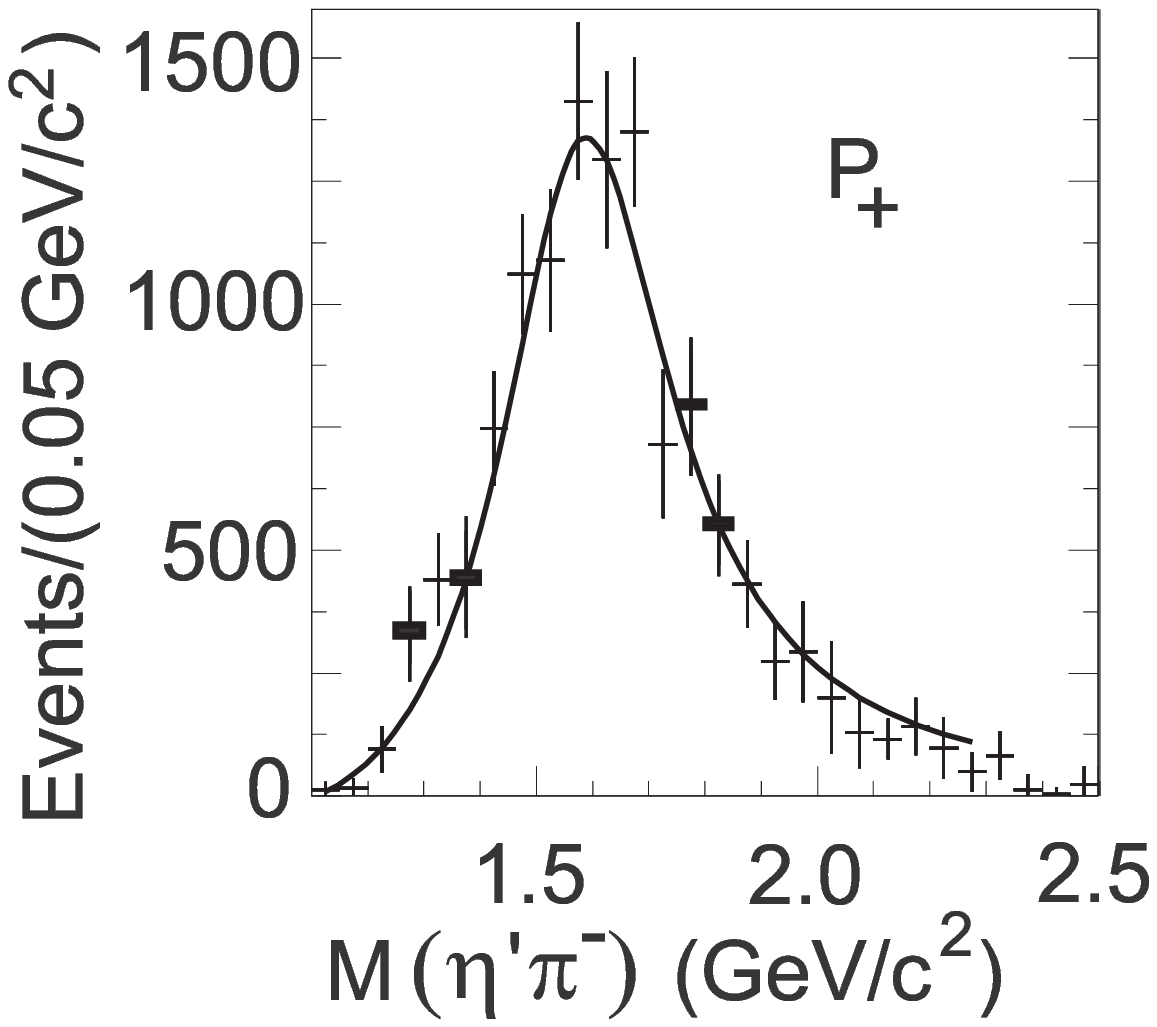,width=.47\linewidth,clip=}
\end{center}
\vspace{-.5cm}\caption{\label{merc}Mercedes-Benz logo (left) and 
number of events (per $50$ MeV bin)
as a function of $\eta'\pi^-$ invariant mass~\protect\cite{e852} (right).}
\end{figure}

(2) {\it Gluonic excitations are qualitatively new}. 
While systems with more degrees of freedom
than quarks are qualitatively new, the most promising
place to study gluonic excitations does not appear to
be hybrid baryons. Glueballs (made from gluons) and hybrid
mesons (a quark and antiquark with a gluonic excitation)
are more promising since they can be $J^{PC}$ exotic, meaning
that there are no mesons in the quark model with these
$J^{PC}$, or there are no local quark--antiquark currents with 
these $J^{PC}$. Hybrid baryons have half--integral $J$ and
no $C$, and there are no $J^P$ exotics: all  $J^P$ can be
constructed for baryons in the quark model.
In fact, a $J^{PC}=1^{-+}$ exotic isovector meson at $1.6$ GeV 
has recently been reported in $\eta'\pi^-$.~\cite{e852} 
In this analysis the exotic partial wave is in fact the 
dominant one. If one were shown the event shape
(Fig.~\ref{merc}) a few decades ago when the $\rho$ was 
discovered, it would be easy to conclude that a new
resonance has been discovered. The exotic meson is currently
thought to be a hybrid meson.

\begin{figure}[t]
\begin{center}
\epsfig{file=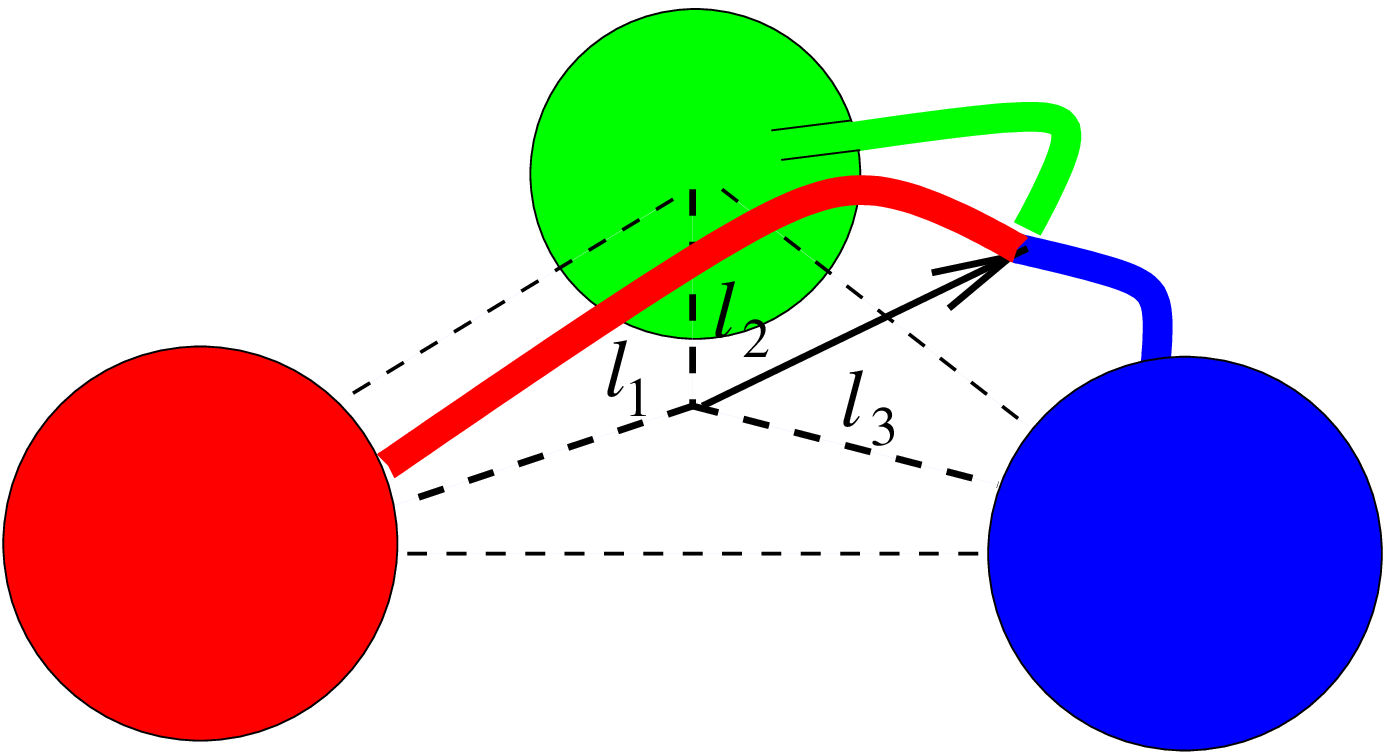,width=.43\linewidth,clip=}
\epsfig{file=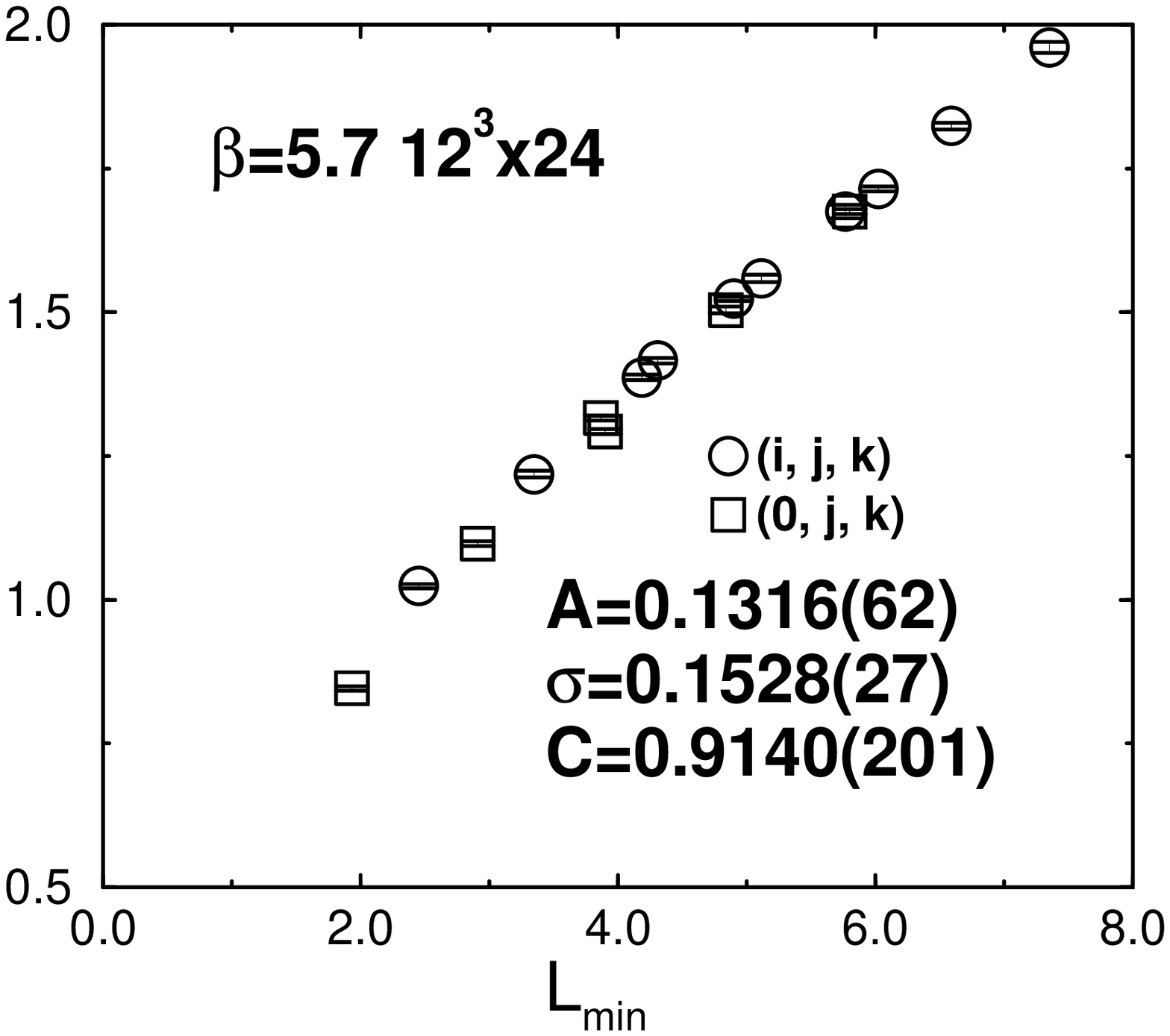,width=.53\linewidth,clip=}
\end{center}
\vspace{-.5cm}\caption{\label{latt}Definition of $l_1,l_2$ and $l_3$ (left) and 
lattice potential~\protect\cite{lat} (right).}
\end{figure}

(3) {\it Hybrid baryons are a test of intuitive pictures of 
Quantum Chromodymanics (QCD)}.
As we will shortly detail, current lattice QCD data indicate that
the interaction between three quarks is indeed a 
$Y$--shaped potential, which is expected to arise from 
three gluon flux--tubes meeting at a junction (see Fig. \ref{latt}). Excitations of
these gluon flux--tubes is expected to be very sensitive
to the way that the flux--tubes connect (in this case in a
$Y$--shape): hence the importance of hybrids for our
intuitive pictures of QCD. The gluon self--interaction 
has no analogue in the Abrikosov--Nielsen--Olesen flux--tubes 
of QED. 

As promised, let us sketch current lattice QCD data. In 
Fig.~\ref{latt} we define the three lengths $l_1,l_2$ and $l_3$ as
the distances from the quarks (the blobs) to the equilibrium
position of the junction of the $Y$--shaped linearly
confining flux--tube system. 
Define ${\mbox L}_{\mbox{min}}=l_1+l_2+l_3$. The quenched lattice
potential for the three quarks as a function of 
${\mbox L}_{\mbox{min}}$ is plotted in Fig.~\ref{latt}.
At large ${\mbox L}_{\mbox{min}}$ the potential is 
proportional to ${\mbox L}_{\mbox{min}}$. The potential
was parameterized by a Coulomb and $Y$--shaped confining term~\cite{lat}

\begin{equation}
{V_{3Q}} = -\mbox{A}\sum_{i<j}\frac{1}{|\br_i-\br_j|}+{\sigma {\mbox L}_{\mbox{min}}}+\mbox{C}
\label{lat}
\end{equation}
and the string tension $\sigma = 0.1528(27)$ GeV$^2$ was found to 
be similar to the value measured between a quark and an antiquark,
albeit somewhat smaller. A possible reason for this is a
cancellation of chromo--electric fields at the junction of the
$Y$--shaped flux--tube.~\cite{simonov} The 
value of the string tension is consistent with the value of 
$0.15$ GeV$^2$ 
extracted from the experimental baryon spectrum by Capstick and 
Isgur.~\cite{capisg} 
It is found that $\chi^2/d.o.f.$ = 3.99 for the $Y$--shaped 
confinement potential versus and 10.9 for pairwise 
confinement, so that $Y$--shaped confinement is clearly
preferred. Another lattice work~\cite{alter} concludes that 
pairwise confinement is preferred over $Y$--shaped confinement,
but imposes the constraint that $\sigma$ for baryons must be
identical to $\sigma$ for mesons, an assumption which is 
unjustified.~\cite{simonov}

\section{What are hybrid baryons?}

Hybrid baryons can be defined in two ways:

(1) {\it Three quarks and a gluon}. 
If states are rigorously expanded in Fock space, 
one can discuss hybrid (three quark -- gluon)
components of such an expansion, which can be accessed in 
large $Q^2$ deep inelastic scattering.~\cite{carlson}
Historically a low--lying hybrid baryon was defined as a three 
quark -- gluon 
composite. However, from the viewpoint of the Lagrangian of 
QCD this definition is non--sensical.
This is because gluons are massless, and hence there is no reason not
to define a hybrid baryon, for example, as a three quark -- two 
gluon 
composite. Neither is one possibility distinguishable from the other,
since strong interactions mix the possibilities. 
Moreover, sometimes the definition becomes
perilous. A case in point is recent work on large $N_c$ hybrid
baryons, where their properties depend critically on the fact that 
the
gluon is in colour octet, and hence the three quarks in colour octet,
so that the entire state is colour singlet.~\cite{pirjol}
The bag model circumvents the objections raised against this 
definition,
since gluons become massive due to their confinement inside 
the bag.~\cite{bag,golowich,carl,ch,duck}

\begin{figure}[t]
\begin{center}\label{adiab}
\vspace{-1.7cm}\epsfig{file=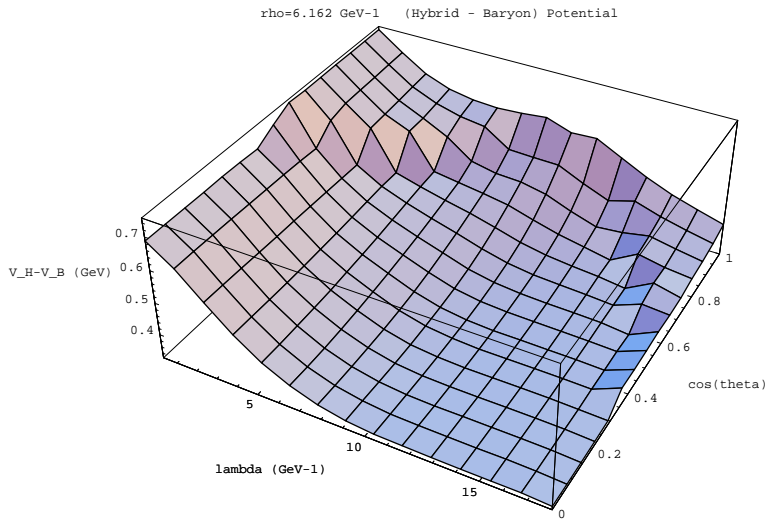,width=0.85\linewidth,clip=}
\end{center}
\vspace{-2cm}\caption{Difference between hybrid and 
conventional baryon adiabatic potentials as a function of the 
quark positions (parametrized in terms of the Jacobi coordinates
$\rho,\lambda$ and $\theta$, with $\rho$ fixed in this 
case.)~\protect\cite{adia}}
\end{figure}

\begin{figure}[t]
\begin{center}
\epsfig{file=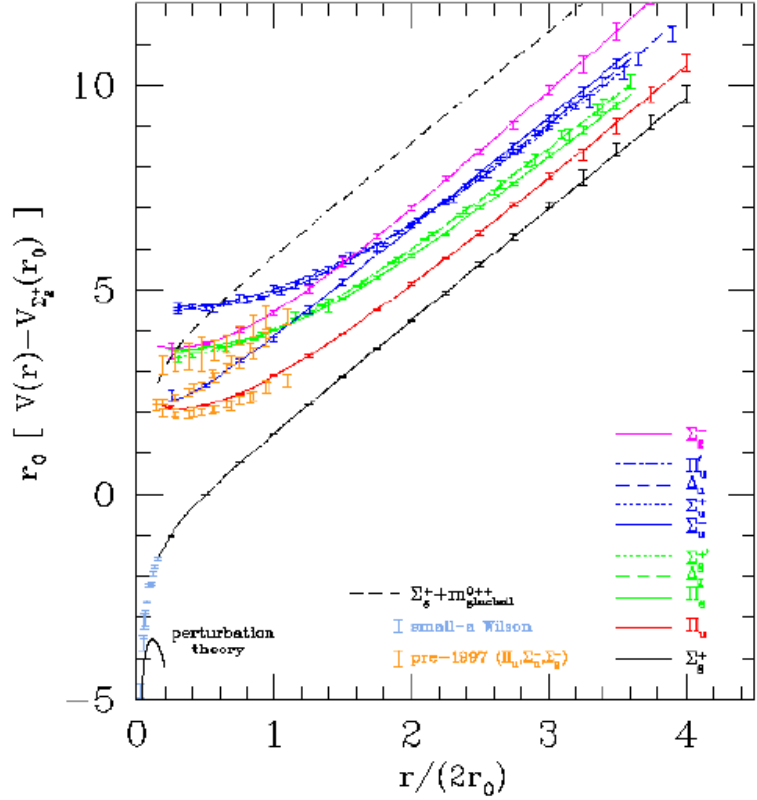,width=0.9\linewidth,clip=}
\end{center}
\vspace{-0.5cm}\caption{\label{morning}Adiabatic potentials as a function of
$Q\bar{Q}$ separation.~\protect\cite{morn}}
\end{figure}

(2) {\it Three quarks moving is an excited adiabatic potential}.
One can always evaluate
the energy of a system of three fixed quarks as a function of the 
three quark positions, called the adiabatic potential.
There is a ground--state adiabatic potential, corresponding
to conventional baryons, and various excited adiabatic potentials,
corresponding to hybrid baryons. 
The three quarks are then allowed to move in the excited
adiabatic potential. 
This can be a perfectly sensible definition
from the viewpoint of QCD.
The caveat is that this definition is only exact for
(1) very heavy quarks (for some potentials), or
(2) for specific simplified dynamics, particularly that of
three non--relativistic quarks moving in a simple harmonic oscillator
potential.~\cite{nstar} In the latter case the definition is
exact even for up and down constituent quarks if one
redefines the adiabatic potential suitably.~\cite{nstar}
For the linear potentials of the flux--tube model is was noted
that ``For light quarks almost all corrections may be incorporated 
into a redefinition of the potentials. Mixing between [new] 
potentials is of the order of 1\%''.~\cite{merlin}
If mixing between the (redefined) adiabatic potentials is this
small one can sensibly talk about hybrids even for light
quarks. The redefinition depends on the quark masses, so that
one would ideally start with very heavy quarks on the original
adiabatic potential, and then gradually move to the quark masses of 
interest by redefining the adiabatic potential. 
An example of an adiabatic surface in the flux--tube model
appears in Fig.~\ref{adiab}. The potential
peaks at small $\rho$ and $\lambda$ and the uneven rim 
corresponds to the
transition from a $Y$--shaped flux--tube to a
two--legged flux--tube (when the angle between two of the
quarks is more than $120^o$).

\section{How are hybrid baryons modelled?}

To answer this question it is best to understand how 
hybrid mesons can be modelled. The quenched lattice QCD hybrid meson 
adiabatic potentials are shown in Fig.~\ref{morning}.
(Note that at large $Q\bar{Q}$ separation,
linear confinement should break down due to $q\bar{q}$ pair 
creation. For low--lying hybrids
this effect can be incorporated as a higher order
effect, i.e. as loop corrections to masses.)
At small $Q\bar{Q}$ separation, 
the adiabatic bag model (where quarks are
stationary) gives a reasonable description of the
lattice data.~\cite{mornbag} At large $Q\bar{Q}$ separations, 
a constituent gluon model (related
to the bag model) is not applicable,~\cite{swanson99} but a
Nambu--Goto string (flux--tube) picture instead~\cite{olsson}.
One hence needs a combined phenomenology of the
``old bag'' model and the 
flux--tube model to model hybrid mesons, and
by implication hybrid baryons. This is a somewhat unhappy
marriage, as the two models describe glue
very differently. The exact way the glue is modelled is 
critical, e.g. in the flux--tube model $Y$--shaped confinement 
for low--lying hybrid baryons has been shown to be 
well-approximated by motion of the junction of the 
$Y$--shape.~\cite{capstick} The dynamics is critically 
dependent of what the nature of the excitation is.

We now summarize model estimates for
masses of hybrid baryons. QCD sum rules estimates the
low--lying $N \frac{1}{2}^+$ hybrid at
$\sim 1.5$ GeV.~\cite{sum} The bag model obtains the
lightest hybrid, $N \frac{1}{2}^+$,
at $\sim {1.55}$ GeV,~\cite{bag,golowich,carl}
between the $N(1440)$ (Roper) 
and $N(1710)$. Higher mass $N$ and $\Delta$ 
hybrids are at $1.5-2.5$ 
GeV.~\cite{bag,golowich,carl} In the flux--tube model the $N$ hybrids
are at $1.87(10)$ GeV and the $\Delta$ hybrids
at $2.08(10)$ GeV.~\cite{capstick}
Hybrids with strangeness have surprisingly low  
mass, particularly a flavour singlet $\Lambda$ at
$\sim 1.65$ GeV in the bag model.~\cite{carl,ch}

\begin{table}[b]
\renewcommand{\tabcolsep}{0.4pc} 
\renewcommand{\arraystretch}{1.2} 
\begin{center}
\caption{\label{ft}Quantum numbers in the flux--tube 
model.~\protect\cite{capstick}}\vspace{.25cm}
\begin{tabular}{|lc|cl|}
\hline
Mass&{Flavour} & {Spin} & $J^{P}$ \\
\hline\hline
Light&{$N$}  & {$\frac{1}{2}$} & $\frac{1}{2}^{+},\;\frac{3}{2}^{+}$ \\
Light&{$N$}  & {{$\frac{1}{2}$}} & {$\frac{1}{2}^{+},\;\frac{3}{2}^{+}$} \\
Heavy&{$\Delta$} & {{$\frac{3}{2}$}} & {$\frac{1}{2}^{+},\;\frac{3}{2}^{+},\;\frac{5}{2}^{+}$} \\
\hline
\end{tabular}\end{center}
\end{table}

What are the quantum numbers of hybrid baryons? The good
quantum numbers are flavour (to the extent that isospin is
a good symmetry), $J$ and $P$. The non-relativistic spin
of the three quarks is also important in (non--relativistic)
models, but is not a good quantum number. 
Table \ref{ft} lists the
quantum numbers of the low--lying hybrids in the flux--tube
model. The total angular momentum $J$ is obtained by 
adding the spin (either 
$\frac{1}{2}$ or $\frac{3}{2}$) to unit orbital angular  
momentum $L$. The four $N$ hybrids are the
lightest, and the five $\Delta$ hybrids heavier, as 
previously remarked. 
Although hybrids contain the quantum numbers of the conventional $N$ 
and $\Delta$, one never obtains 
a full multiplet of conventional baryons like this in the quark 
model, even for excited conventional baryons. It is, however,
possible to have $L^P=1^+$ for conventional baryons, as is the
case for hybrids.
The quantum numbers of hybrids in the bag model is shown in 
Table~\ref{bb}. These differ from the flux--tube model in the
last two rows: The spin $\frac{1}{2}$ and  $\frac{3}{2}$
is exchanged, with corresponding changes in $J$. 
In the bag model the $N\frac{1}{2}^+$ is the lightest,
then the $N\frac{1}{2}^+$, $N\frac{3}{2}^+$
and $N\frac{3}{2}^+$, then the $\Delta\frac{1}{2}^+$ and
$\Delta\frac{3}{2}^+$ with $N\frac{5}{2}^+$ the heaviest.
The bag model has the same number of low--lying
states as in flux--tube model.
When good quantum numbers are considered, 
the only difference is that the $J^P=\frac{5}{2}^+$ state is 
a $\Delta$ in the flux--tube model and a $N$
in the bag model. In both models this is one of the highest
lying states, so that the low--lying hybrids are identical.
In fact, $N\frac{1}{2}^+$ is amongst the lightest hybrids
in both models, and, as previously noted, is light in
QCD sum rules as well. For hybrids with strangeness,
the $\Lambda\frac{1}{2}^+$ and $\Lambda\frac{3}{2}^+$
were predicted in the bag model, with the $\Lambda\frac{1}{2}^+$
the lighter state.~\cite{carl,ch}

\begin{table}[t]
\renewcommand{\tabcolsep}{0.4pc} 
\renewcommand{\arraystretch}{1.2} 
\begin{center}
\caption{\label{bb}Quantum numbers in the bag
model.~\protect\cite{bag,golowich,carl}}\vspace{.25cm}
\begin{tabular}{|c|cl|}
\hline
{Flavour} & {Spin} & $J^{P}$ \\
\hline\hline
{$N$}  &{$\frac{1}{2}$} & $\frac{1}{2}^{+},\;\frac{3}{2}^{+}$ \\
{$N$}  &{{$\frac{3}{2}$}} & {$\frac{1}{2}^{+},\;\frac{3}{2}^{+},\;\frac{5}{2}^{+}$} \\
{$\Delta$} & {{$\frac{1}{2}$}} & {$\frac{1}{2}^{+},\;\frac{3}{2}^{+}$} \\
\hline
\end{tabular}\end{center}
\end{table}

Recently, masses and quantum numbers of hybrid baryons have been
reported in a dispersion relation technique.~\cite{kochkin}

\section{How does one find hybrid baryons?}

There are currently two approaches:

\begin{itemize}

\item Comparison with models (which are themselves callibrated 
against lattice QCD {\it and} experiment on glueballs 
and hybrid mesons).

\item Comparison to generic expectations for glue--rich hadrons.

\end{itemize}

Hybrid baryons can be found by

\begin{figure}[t]
\begin{center}\label{spec}
\epsfig{file=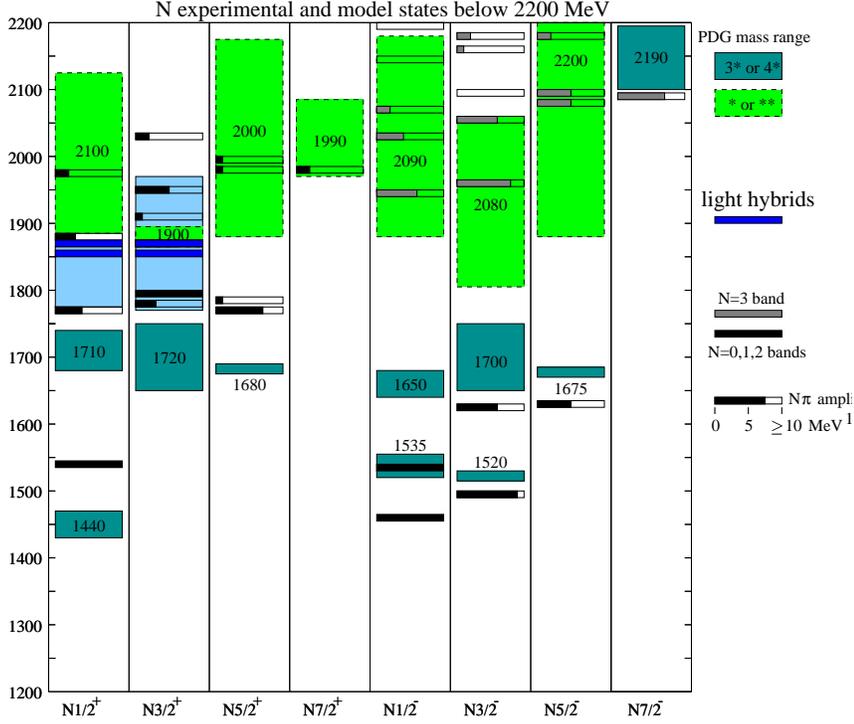,width=0.98\linewidth,clip=}
\end{center}
\caption{Model (conventional and hybrid) and experimental 
$N$ baryons. For each set of quantum numbers, the mass spectrum 
is indicated (in MeV). The thin bars indicate quark model
predictions for conventional baryon masses.~\protect\cite{capisg}
The four fully filled thin bars around $1870$ MeV are the flux--tube model
hybrids.~\protect\cite{capstick}
The large rectangular blocks are the mass ranges of known
experimental states.}
\end{figure}

(1) {\it Observing more states than the conventional baryons.}
This approach is very difficult in practice, and has only
proved useful in the $J^{PC}=0^{++}$ (scalar) 
isoscalar meson sector, which has led to the identification of
a gluonic excitation: the glueball.
To see how difficult this approach is for hybrid baryons,
look at Fig.~\ref{spec}. Nowhere in the spectrum does one
observe an excess of quark model (conventional) baryons
above those known experimentally. There is hence no need
to posit hybrids. 

Another possible way to identify hybrids in the spectrum is to compare
the behaviour of experimental states relative to properties conventional
baryons are {\it known} to have. Examples of such properties
is the hypothesis that the high--lying conventional baryons
occupy chiral multiplets~\cite{glozman} or that conventional
baryons follow Regge trajectories~\cite{klempt}.

{\it (2) Diffractive $\gamma N$ and $\pi N$ production.}
The detection of the hybrid meson candidate $\pi(1800)$ 
in diffractive
$\pi N$ collisions by VES~\cite{ves} may indicate that hybrid mesons
are producted abundantly via meson--pomeron fusion. If this is the 
case,
one expects significant production of hybrid baryons via 
baryon--pomeron fusion, i.e. production in 
diffractive $\gamma N$ and $\pi N$ collisions.

{\it (3) Production in $\psi$ decays.} Na\"{\i}ve expectations
are that the gluon--rich environment of $\psi$ decays should lead to 
dominant production of glueballs, but also signifant production of
hybrid mesons and baryons. The large branching ratios~\cite{pdg00}
$Br(\psi\rightarrow p\bar{p}{\omega},\;  p\bar{p}\eta') \sim
10^{-3}$  may indicate hybrid baryons. Recently a 
$J^P=\frac{1}{2}^+$ $2\sigma$ peak at mass $1834^{+46}_{-55}$ MeV
was seen in $\Psi\rightarrow p\bar{p}\eta$.~\cite{zou}

{\it (4) Production in $p\bar{p}$ annihilation.} The fact that the
scalar glueball is strongly produced in this process, although
not dominantly, may make it a promising production process.

{\it (5) Studying hybrid baryon decays.} Except for a QCD sum rule 
motivated suggestion that the
$N\frac{1}{2}^+$ hybrid baryon has appreciable\footnote{
They find that
$\Gamma_\sigma/\Gamma_{tot} =$ only 10\% (modulo phase space), 
consistent with the Roper, although  
$\Gamma_\sigma$ is still much larger than for other resonances.} 
decay to $N\sigma$,~\cite{li} 
and a bag model calculation which predicts
a $N\frac{3}{2}^+$ with large $\pi\Delta$ decay and minute
$\pi N$ decay,~\cite{duck} no decay calculations
have been performed. However, decay of hybrid baryons to 
$N\rho$ and $N\omega$ is {\it a priori} interesting since it isolates
states in the correct mass region, without contamination 
from lower--lying
conventional baryons. 

{\it (6) Electroproduction.} In the flux--tube model,
which is an adiabatic picture of a hybrid baryon, there
is the qualitative conclusion that
``$e p\rightarrow e X$ should produce ordinary $N^{\ast}$'s and
hybrid 
baryons with comparable cross--sections''.~\cite{bar}
However, the conclusions obtained from the three 
quark -- gluon picture
of a hybrid baryon is different. For large $Q^2$ electroproduction,
the $Q^2$ dependence of the amplitudes is summarized in Table 
\ref{qsq}.
Since the photon has both a transverse and 
longitudinal component, the amplitude for a conventional baryon 
is expected to dominate that of the hybrid baryon as $Q^2$ becomes 
large.~\cite{carlson}
For small $Q^2$ the conclusion agrees with the large $Q^2$ result
for transverse photons, but is more dramatic for longitudinal photons:
the amplitude vanishes.~\cite{volker,tb}
It has accordingly been concluded that the ({radially } excited) conventional 
baryon is dominantly electroproduced (depending on the details of
the calculation), 
with the {hybrid} baryon subdominant relative 
to the resonances $S_{11}(1535), D_{13}(1520)$ and $\Delta$ as 
$Q^2$ increases.~\cite{volker} The $Q^2$ dependence of the 
electroproduction of a resonance can be measured at 
Jefferson Lab Halls B and C and an
energy upgraded Jefferson Lab. A hybrid baryon is expected to behave
different from nearby conventional baryons as a function of $Q^2$. One
needs to perform partial wave analysis at different $Q^2$. For
large $Q^2$ cross--sections are small, which would make this way of
distinguishing conventional from hybrid baryons challenging.

\begin{table}
\renewcommand{\tabcolsep}{0.2pc} 
\renewcommand{\arraystretch}{0.9} 
\begin{center}
\caption{\label{qsq}$Q^2$ dependence of amplitudes for the 
electroproduction of
conventional or hybrid baryons with transverse or longitudinal 
photons, valid at large $Q^2$.~\protect\cite{carlson}
\protect\label{carlsontable}}\vspace{.25cm}
\begin{tabular}{|c||c|c|}
\hline
 & {Conventional} & { Hybrid} \\
\hline\hline
{ Transverse}  &  $1/Q^3$ & $1/Q^5$ \\
{ Longitudinal}&  $1/Q^4$ & $1/Q^4$ \\
\hline
\end{tabular}\end{center}
\end{table}

\vspace{.3cm}

Various characteristics of the $N(1440)$ (Roper) have been 
argued to be consistent with its being dominantly a hybrid baryon:
Its mass is consistent with bag model~\cite{golowich} and 
QCD sum rule~\cite{sum} estimates, it is suppressed in
large $Q^2$ electroproduction~\cite{volker} and the width
ratio $\Gamma_\sigma / \Gamma_{tot}$ is consistent with QCD sum
rule decay calculations~\cite{li}.

The $\Lambda(1405)\; (J=\frac{1}{2})$ and $\Lambda(1520)
\; (J=\frac{3}{2})$
have recently been proposed to be hybrid
baryons based on the idea that this hypothesis solves a $J$
ordering problem in this system: for hybrid baryons the ordering
in a constituent gluon model is as observed,
while it is opposite in (most) quark models.~\cite{farrar}

We highlight current experimental searches for hybrid baryons.
Photo-- and electroproduction 
efforts at Halls B and C at Jefferson Lab (Newport News)
can isolate hybrid baryons
in ${\gamma} N\rightarrow \mbox{hybrid} \rightarrow 
({\rho},{\omega},{\eta})N$ at masses less than $2.2$ GeV. 
As previously remarked, the higher mass decay channels are of
most interest. Similar searches at the
Crystal Barrel and SAPHIR detectors at ELSA (Bonn) is under way.
Particularly, at Crystal Barrel hybrid baryons is planned to be
isolated in ${\gamma} N\rightarrow \mbox{hybrid} \rightarrow 
({\eta},\pi^0)\; S_{11}(1535)\rightarrow  ({\eta},\pi^0)\; 
{\eta} N$. Hybrid mesons in flux--tube model 
decay strongly to $P+S$--wave mesons and not to
$S+S$--wave mesons. If this is also true for hybrid baryons,
decay to a $P$--wave baryon ($S_{11}(1535)$) and an 
$S$--wave meson (${\eta}$ or $\pi^0$) should be prominent.
In $\Psi$ production hybrid baryons are searched for at
BES at BEPC. Searches for hybrid baryons in 
$\Psi\rightarrow \mbox{hybrid} \;\bar{p} 
\rightarrow p\bar{p}\; ({\eta},\pi^0)$  have been 
undertaken.~\cite{zou}

\section*{Acknowledgments}
This research is supported by the Department of Energy under contract
W-7405-ENG-36.

\end{document}